# Stochastic dynamics and the dynamic phase transition in thin ferromagnetic films

Hyunbum Jang[1], Malcolm J. Grimson[2], and Thomas B. Woolf[1]

[1]*Department of Physiology, Johns Hopkins University, Baltimore, MD 21205*
[2]*Department of Physics, University of Auckland, Auckland, New Zealand*

The dynamic phase behavior of a classical Heisenberg spin system with a bilinear exchange anisotropy $\Lambda$ in a planar thin film geometry has been investigated by Monte Carlo simulations using different forms for the stochastic dynamics. In simulations of the dynamic phase transition (DPT) in films subject to a pulsed oscillatory external field with competing surface fields, both Glauber and Metropolis dynamics show a continuous DPT. Furthermore the field amplitude dependence of the DPT is qualitatively similar for both Glauber and Metropolis dynamics. However, the temperature dependence of the DPT is markedly different with the DPT being much sharper for Glauber dynamics. The difference arises from a decoupling of the surface and bulk responses of the film near the dynamic phase transition with Metropolis dynamics that is not evident for Glauber dynamics.

## I. INTRODUCTION

In Monte Carlo simulations the algorithm incorporates a stochastic dynamics that provides a rule whereby the system changes from one state to another. There are many different possible types of stochastic dynamics that can involve either single or many particle moves and thereby be either local or non-local in character. In equilibrium Monte Carlo simulations a number of different dynamics can lead to the same Boltzmann distribution of states once the simulation has reached equilibrium. The conditions of ergodicity and detailed balance are sufficient to ensure that the equilibrium distribution of states sampled by the algorithm is the correct Boltzmann distribution [1]. Thus one is free to choose any algorithm that obeys ergodicity and detailed balance and one should get the same result in an equilibrium Monte Carlo simulation. So the computationally most efficient algorithm is usually selected.

However, the fundamental difficulty that makes nonequilibrium Monte Carlo simulations harder than their equilibrium counterparts is that there is a limited freedom in choosing the dynamics of the Monte Carlo algorithm [1]. The conditions of ergodicity and detailed balance say nothing about the way in which the system comes to equilibrium and different choices for the stochastic dynamics will give rise to different results. Thus the dynamic must be chosen on physical grounds rather than simple computational efficiency and for cluster algorithms the relation of the Monte Carlo process to a realistic dynamical process is unclear [2].

In some cases, when simulating a real material, it is possible to use our understanding of that material to estimate the correct form for the stochastic dynamics, so as to make the Monte Carlo algorithm mimic the true system as closely as possible. However in other cases the detailed form of the dynamics is not *a priori* clear and macroscopic properties must be used to make some inference as to the form of the stochastic dynamics. Thus it is important to understand the non-equilibrium statistical mechanics of model systems with well-characterised stochastic dynamics.

In recent papers [3-5], Rikvold and Kolesik have shown that the interface structure and velocity in a kinetic Ising ferromagnet driven by an applied field depends strongly on the details of the stochastic dynamics. Here we shall investigate the role of the type of stochastic dynamics on the dynamic phase transition (DPT) observed in the thin ferromagnetic film with competing surface fields where the dynamic variation of the magnetization in the film is the result of interface motion within the film [11-13]. Section II contains a description of the model under investigation and the dynamic Monte Carlo method used. Sections III present the temperature and field amplitude dependence of the DPT, comparing and contrasting the results for the different forms of stochastic dynamics. The paper concludes with a discussion.

## II. MODEL

The system under consideration here is a three dimensional thin planar film of finite thickness $D$ with competing surface fields subject to a time dependent oscillatory external field $H(t)$ with Hamiltonian

$$H(t) = H_0 - h\left(\sum_{i \in \text{surface} 1} S_i^z - \sum_{i \in \text{surface} D} S_i^z\right) - H(t)\sum_i S_i^z \quad . \quad (1)$$

The competing surface fields are characterized by a magnitude $h$ and $H(t)$ is taken to have a pulsed form with

$$H(t) = \begin{cases} -H_0, & \frac{2(k-1)\pi}{\omega} < t \leq \frac{(2k-1)\pi}{\omega} \\ H_0, & \frac{(2k-1)\pi}{\omega} < t \leq \frac{2k\pi}{\omega} \end{cases}, \quad (2)$$

where $H_0$ is the amplitude, $\omega$ is the angular frequency and $k$ ($k = 1, 2, 3, \ldots$) is an integer representing the number of periods of the pulsed oscillatory. The anisotropic classical Heisenberg model is defined by

$$H_0 = -J\sum_{\langle i,j \rangle}\left((1-\Lambda)(S_i^x S_j^x + S_i^y S_j^y) + S_i^z S_j^z\right), \quad (3)$$

where $\mathbf{S}_i = (S_i^x, S_i^y, S_i^z)$ is a unit vector representing the $i$th spin and the notation $\langle i,j \rangle$ indicates that the sum is restricted to nearest-neighbor pairs of spins. $J$ is a coupling constant characterizing the magnitude of the exchange interaction and for ferromagnets $J > 0$. $\Lambda$ characterizes the strength of the bilinear exchange anisotropy. In the isotropic limit, $\Lambda = 0$, the model reduces to the familiar classical Heisenberg model, while for $\Lambda = 1$, the Hamiltonian becomes Ising-like [6].

The model film is a simple lattice of size $L \times L \times D$, in units of the lattice spacing. Periodic boundary conditions are applied in the $x$ and $y$ directions. Free boundary conditions are applied in the $z$ direction that is of finite thickness $D$. The system is subject to competing applied surface fields of magnitude $h = -0.55$ in layers $n = 1$ and $n = D$ of the film. A film thickness $D = 12$ was used throughout. This value corresponding to the crossover regime between wall and bulk dominated behavior for thin Ising films [7]. In thinner films it is difficult to distinguish between "interface" and "bulk" phases in the film, since all layers of the film feel the effect of the competing surface fields rather strongly. While for thicker films the surfaces of the film only interact close to the bulk critical point. The results reported here are for lattices of size $L = 32$, but no significant differences were found for lattices with $L = 64$ and 128 at non-critical values of $H_0$, $\omega$ and $T$.

Monte Carlo simulations were performed using a random spin update scheme. Trial configurations were generated by the rotation of a randomly selected spin through a random angular displacement about one of the $x, y, z$ axes chosen at random [8,9]. A sequence of size $L \times L \times D$ trials comprises one Monte Carlo step per spin (MCSS), the unit of time in our simulations. The period of the pulsed oscillatory external field is given by product $R_{FS} \times N$, where $R_{FS}$ is the field sweep rate [10] and $N$ is a number of MCSS. The applied oscillatory field $H(t)$ being updated after every MCSS according to Eq. (2). The simulations reported here were performed for a value of $R_{FS} = 1$ with $N = 240$. In all of the simulations, the initial spin configuration was a ferromagnetically ordered state of the spins with $S_i = +1$ for all $i$.

The order parameter for the DPT is the period averaged magnetization over a complete cycle of the pulsed field, $Q$, defined by

$$Q = \frac{\omega}{2\pi}\oint M_z(t)\,dt, \quad (4)$$

where the $z$-component of the magnetization for the film is

$$M_z(t) = \frac{1}{D}\sum_{n=1}^{D} M_n^z(t), \quad (5)$$

with

$$M_n^z(t) = \frac{1}{L^2}\sum S_i^z(t) \quad (6)$$

being the $z$-component of the magnetization for the $n$th layer of the film. The system exhibits a dynamically ordered phase with $|Q| > 0$ and a dynamically disordered phase with $Q = 0$. The period averaged magnetization for the $n$th layer of the film is given by

$$Q_n = \frac{\omega}{2\pi}\oint M_n^z(t)\,dt. \quad (7)$$

### III. GLAUBER VS. METROPOLIS DYNAMICS

Previous studies [11-13] of hysteresis and the dynamic phase transition (DPT) in anisotropic Heisenberg ferromagnets driven by an oscillatory applied external field have used a dynamic Monte Carlo simulation method with Metropolis dynamics. Metropolis dynamics is defined by the transition probability

$$W_M(s_i \rightarrow s_i') = \min[1, e^{-\beta \Delta E}], \quad (8)$$

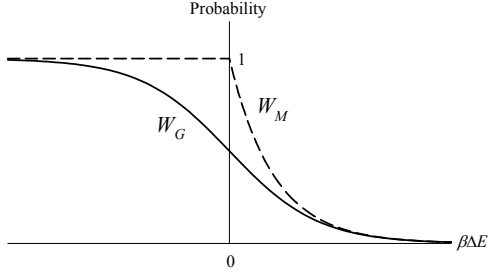

**FIG. 1** The acceptance probability for a trial spin rotation of energy $\Delta E$ for Metropolis (dotted line) and Glauber (solid line) dynamics.

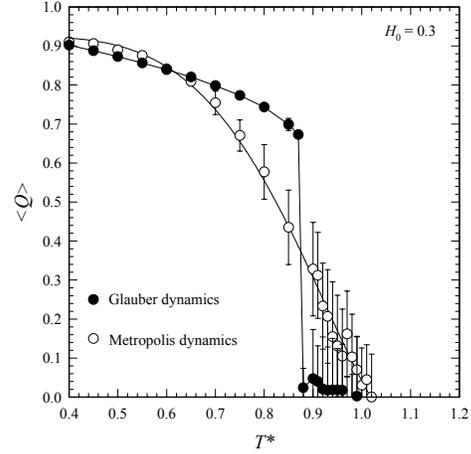

**FIG. 2** Mean period averaged magnetization $\langle Q \rangle$ as a function of the temperature $T^*$ with a fixed value of the pulsed oscillatory external field amplitude of $H_0 = 0.3$ for Glauber (solid circles) and Metropolis (open circles) dynamics.

where $\Delta E$ is the total energy change associated with the trial rotation of the $i$th spin $S_i \rightarrow S_i'$. In all cases the unit of time in the simulation is a Monte Carlo step per spin (MCSS). However in the Monte Carlo method no physical time is associated with each trial configuration. One MCSS simply corresponds to a series of random modifications of all the degrees of freedom of the system. But if the time rate by which a real system can modify all of its degrees of freedom is known by some independent argument, then the number of MCSS can be converted into a real time unit [14,15]. However in doing this one must be sure that the form of the stochastic dynamics used in the Monte Carlo simulation is appropriate, particularly if the dynamic response of the system depends strongly on the details of the stochastic dynamics.

Glauber dynamics [1] is defined by the transition probability

$$W_G(s_i \rightarrow s_i') = \frac{e^{-\beta \Delta E}}{1 + e^{-\beta \Delta E}} \quad . \tag{9}$$

Both Glauber and Metropolis dynamics obey ergodicity and detailed balance. The two are compared graphically in Fig.1. If $|\beta \Delta E| \gg 1$, then $W_M \approx W_G$. The only significant difference occurs for $|\beta \Delta E| \ll 1$, when $W_M > W_G$. Thus Metropolis dynamics is always more likely to accept a trial spin rotation that involves a small change in energy, particularly when trial spin rotation lowers the energy. However, it is important to note that the transition probability for both Metropolis and Glauber dynamics depends directly on $\Delta E$, the total energy change associated with the transition. Thus the effect of the applied fields tends to dominate the transition rate such that it is near unity for spin rotations that bring a spin into a parallel alignment with the net applied field and near zero for spin rotations in the opposite direction irrespective of the change in the interaction energy between the spins.

**A. Temperature dependence of the DPT**

Figure 2 shows the mean period averaged magnetization $\langle Q \rangle$ as a function of the reduced temperature, $T^* = k_B T/J$, for a pulsed oscillatory field amplitude $H_0 = 0.3$. The quantity is determined from a sequence of full cycles of the oscillatory field with initial transients discarded. The error bars in the figure correspond to a standard deviation in the measured values and are only visible when they exceed the size of the symbol. Lines joining the symbols in the figure are solely to guide the eye.

The figure shows a change in the dynamic order parameter for Glauber (solid circles) and Metropolis (open circles) dynamics as $T^*$ increases. The low temperature state with $\langle Q \rangle \neq 0$ corresponding to a dynamically ordered phase, while at high temperatures a dynamically disordered phase with $\langle Q \rangle = 0$ is observed. The results for the two types of stochastic dynamics are the same at low temperatures ($T^* < 0.7$) and higher temperatures above the DPT ($T^* > 1.0$). However, for intermediate temperatures in the vicinity of the DPT, the form of $\langle Q \rangle$ as a function of $T^*$ for the two types of dynamics is different. The DPT for Metropolis dynamics appears to be continuous with a steady decrease in the dynamic order parameter as the DPT is approached. Note that fluctuations in the dynamic order parameter close to the DPT are large. Furthermore the fluctuations in $\langle Q \rangle$ for the Metropolis dynamics increase steadily with $T^*$ as the DPT is approached. This is in marked contrast to the same system with Glauber dynamics. The fluctuations in $\langle Q \rangle$ for Glauber dynamics remain small with increasing $T^*$ in the dynamically ordered phase, but following the sharp decrease in $\langle Q \rangle$ at $T^* = 0.88$, there is a marked increase in the size of the fluctuations in $\langle Q \rangle$. The most striking aspect of the figure is that while the qualitative form of the DPT is markedly different between Glauber and Metropolis dynamics, the locations of the DPT for Glauber and Metropolis dynamics are not so very different.

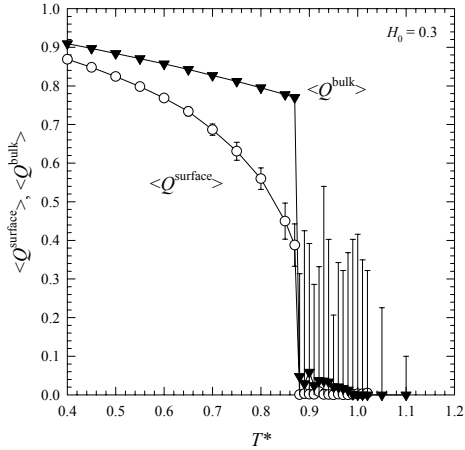

**FIG. 3** Mean period averaged magnetizations of two surfaces $\langle Q^{\text{surface}} \rangle$ and of two central layers $\langle Q^{\text{bulk}} \rangle$ for Glauber dynamics as a function of the temperature $T^*$ with a fixed value of the pulsed oscillatory external field amplitude of $H_0 = 0.3$.

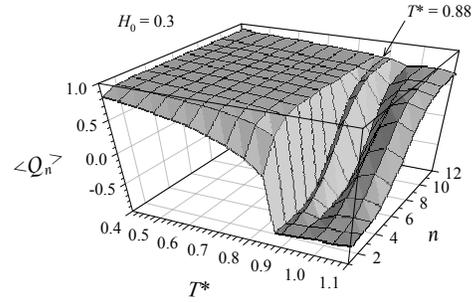

**FIG. 4** Mean period averaged magnetizations for the $n$th layer, $\langle Q_n \rangle$, across the whole film for Glauber dynamics as a function of $T^*$ with a fixed value of the pulsed oscillatory external field amplitude of $H_0 = 0.3$.

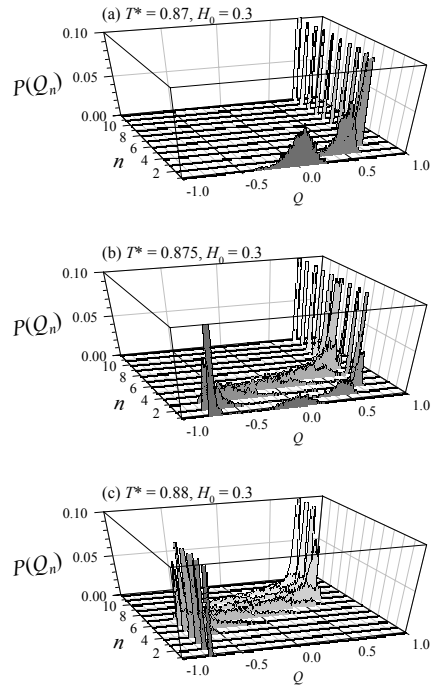

**FIG. 5** Distributions of the layer order parameter $P(Q_n)$ for Glauber dynamics with a fixed value of the pulsed oscillatory external field amplitude of $H_0 = 0.3$ at temperatures $T^* =$ (a) 0.87, (b) 0.875, and (c) 0.88.

Large fluctuations in $Q$ close to the DPT arise from competition between the static surface fields and the pulsed oscillatory external field in the system. To isolate the surface effects, surface order parameter for the film, $Q^{\text{surface}}$, and bulk order parameter for the film, $Q^{\text{bulk}}$, are investigated [12]. The quantities, $Q^{\text{surface}}$ and $Q^{\text{bulk}}$ are simply the mean of the period averaged magnetizations of the two surface and two central layers respectively. Fig. 3 shows $\langle Q^{\text{surface}} \rangle$ and $\langle Q^{\text{bulk}} \rangle$ for Glauber dynamics as a function of $T^*$ for $H_0 = 0.3$. It is immediately clear from the figure that the DPT in the surface layers occurs at a critical temperature of $T_{\text{cd}}^*(\text{surface}) = 0.88$, the temperature at which both $\langle Q \rangle$ and $\langle Q^{\text{bulk}} \rangle$ undergo a sharp decrease. This is markedly in contrast to the same system subject to Metropolis dynamics [12], where the critical temperature for the DPT in the surface layers is distinctly different from that for the DPT in the bulk of the film.

Further information on the form of the DPT follows from mean period averaged layer magnetization $\langle Q_n \rangle$ across the film. Fig. 4 shows the temperature dependence of the order parameter for the $n$th layer, $Q_n$, across the whole film for $H_0 = 0.3$. The figure shows that the DPT in each layer of the film occurs almost at the same temperature. The shape of $\langle Q_n \rangle$ for Glauber dynamics is notably different from the corresponding result for Metropolis dynamics [12] in the bulk of the film. The DPT with Metropolis dynamics is clearly continuous, but this is not clear for the system with Glauber dynamics where the DPT is very sharp.

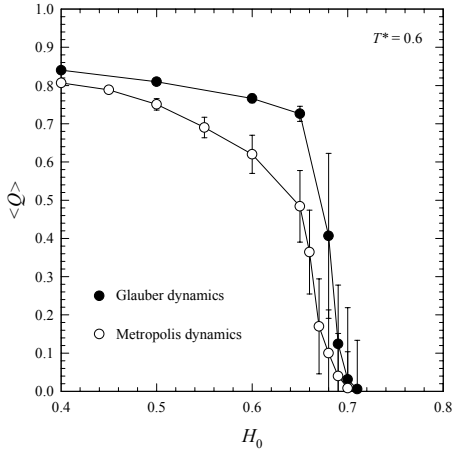

**FIG. 6** Mean period averaged magnetization $\langle Q \rangle$ as a function of the pulsed oscillatory external field amplitude $H_0$ at a fixed value of the temperature of $T^* = 0.6$ for Glauber (solid circles) and Metropolis (open circles) dynamics.

In order to verify the nature of the DPT for Glauber dynamics as a function of $T^*$, the order parameter distributions for the $n$th layer, $P(Q_n)$, across the whole film are obtained. Fig. 5 shows $P(Q_n)$ for $H_0 = 0.3$ at (a) $T^* = 0.87$, (b) $T^* = 0.875$, and (c) $T^* = 0.88$. In the dynamically ordered phase at $T^* = 0.87$, below the DPT, the order parameter distributions for each layer $P(Q_n)$ display a single peak at $+Q$. In the state close to the DPT in the dynamically disordered phase at $T^* = 0.88$, each $P(Q_n)$ shows a single sharp peak located at $\pm Q$. Close to the transition at $T^* = 0.875$, $P(Q_n)$ has a two peak structure in some layers. This shows that the DPT is continuous, because none of the layers in the figure involves a three peak structure with peaks at $\pm Q$ and $Q = 0$. All the layers show a one or two peak structure. The single peak is a result of the surface fields hindering magnetization reversal.

**B. Field amplitude dependence of the DPT**

Fig. 6 shows the mean period averaged magnetization, $\langle Q \rangle$, as a function of the pulsed oscillatory external field amplitude, $H_0$, at a fixed temperatures of $T^* = 0.6$. It is immediately apparent from the figure that the qualitative form of $\langle Q \rangle$ for Glauber dynamics (solid circles) is very similar to that for Metropolis dynamics (open circles) as a function of $H_0$. For both types of dynamics at a fixed temperature the DPT is clearly continuous and $\langle Q \rangle$ vanishes at a value of $H_0 \approx 0.71$.

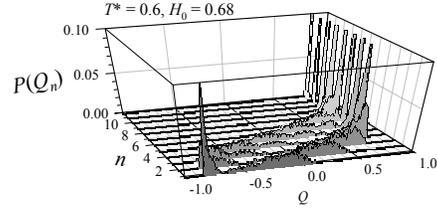

**FIG. 7** Distributions of the layer order parameter $P(Q_n)$ for Glauber dynamics with a fixed temperatures $T^* = 0.6$ for the pulsed oscillatory external field amplitude of $H_0 = 0.68$.

The continuous nature of the DPT for Glauber dynamics as a function of $H_0$ is confirmed in Fig. 7. Close to the transition, $H_0 = 0.68$ at a fixed temperatures of $T^* = 0.6$, $P(Q_n)$ for all the layers of the show a one or two peak structure. There is no evidence for a three peak structure with peaks at $\pm Q$ and $Q = 0$ in any layer of the film.

**IV. CONCLUSIONS**

The different form of the dynamic phase transition (DPT) for the system with Glauber and Metropolis dynamics results the different values for the transition probabilities of the two dynamics in trial spin rotations that involve only small changes in the total energy. In Metropolis dynamics, a trial single spin rotation that results in a reduction of the total energy is accepted. But in Glauber dynamics there is a (small) probability that a lower energy trial configuration will be rejected. Thus high-energy reverse magnetization states can persist to higher temperatures in the film with Glauber dynamics than with Metropolis dynamics. As a result, dynamically ordered states can persist to higher temperatures in systems with Glauber dynamics.

For a given $H_0$, the apparent sudden change in the dynamic order parameter for the film at the DPT with Glauber dynamics is a result of the change in the layer dynamic order parameter occurring at the same temperature for all the layers of the film. In contrast, with Metropolis dynamics [12] the system shows a DPT in the surface layers of the film that occurs at a lower temperature to the DPT in the bulk of the film. This decoupling of the surface and bulk responses of the film to the applied oscillatory field gives rise to a mixed state of the film in which dynamically ordered surfaces coexist with a dynamically disordered bulk. This mixed state of the film persists over a range of temperature and gives rise to an extended region of large fluctuations of the dynamic order parameter for the film. However, while the DPT for the film is much sharper for

Glauber dynamics than with Metropolis dynamics, the DPT is continuous in both cases.

The choice of the transition probability for Glauber dynamics has a physical origin in the interaction of the spin with a heat bath. Whereas, in Metropolis dynamics the transition probability has a mathematical origin, being generated simply from the Metropolis criterion for equilibrium Monte Carlo simulations. However, in the limits of large negative or positive energy changes in the trial spin rotation, the transition probabilities are the same. Thus the two are closely related and the differences only occur for small (negative or positive) energy changes. This work has shown that both types of dynamics give continuous DPTs at similar locations. However, the mixed state that is observed over an extended temperature range near the DPT with Metropolis dynamics is not observed for Glauber dynamics. Thus caution is required in choosing the form of the stochastic dynamics in nonequilibrium Monte Carlo simulations to ensure that the physics of the system is being correctly modeled.

[1] M. E. J. Newman and G. T. Barkema, *Monte Carlo methods in statistical physics*, (Clarendon, Oxford, 1999).
[2] U. W. Nowak, *Annual Reviews of Computational Physics IX*, ed. D. Stauffer, (World Scientific, Singapore, 2001).
[3] P. A. Rikvold and M. Kolesik, J. Phys. A **35**, L117 (2002).
[4] P. A. Rikvold and M. Kolesik, Phys. Rev. E **66**, 066116 (2002).
[5] P. A. Rikvold and M. Kolesik, Phys.Rev. E **67**, 066113 (2003)
[6] K. Binder and D. P. Landau, Phys. Rev. B **13**, 1140 (1976).
[7] K. Binder, D. P. Landau and A. M. Ferrenberg, Phys. Rev. E **51**, 2823 (1995).
[8] R. E. Watson, M. Blume and G. H. Vineyard, Phys. Rev **181**, 811 (1969).
[9] J. R. Barker and R. O. Watts, Chem. Phys. Lett. **3**, 144 (1969).
[10] L. C. Sampio, M. P. de Albuquerque and F. S. de Menezes, Phys. Rev. B **54**, 6465 (1996).
[11] H. Jang and M. J. Grimson, Phys. Rev. E **63**, 066119 (2001).
[12] H. Jang, M. J. Grimson and C. K. Hall, Phys. Rev. B **67**, 094411 (2003).
[13] H. Jang, M. J. Grimson and C. K. Hall, Phys. Rev. E **68**, 046115 (2003).
[14] U. Nowak, R. W. Chantrell and E. C. Kennedy, Phys. Rev. Lett. **84**, 163 (1999).
[15] D. Hinzke and U. Nowak, Phys.Rev. B **61**, 6734 (2000).